# Towards a 6G embedding sustainability

Esteban Selva[1], Azeddine Gati[1], Marie-Hélène Hamon[2]
Orange Innovation Networks
Paris[1], Rennes[2], France
{esteban.selva, azeddine.gati, mhelene.hamon}@orange.com

Bahare Masood Khorsandi[1], Stefan Wunderer[2]
Nokia Strategy & Technology[1]
Nokia Mobile Networks[2]
Munich[1], Ulm[2], Germany
{bahare.masood_khorsandi, stefan.wunderer}@nokia.com

Serge Bories
Univ. Grenoble Alpes, CEA, LETI, Grenoble, France.
serge.bories@cea.fr

Tommy Svensson
Dept. of Electrical Engineering
Chalmers University of Technology
Gothenburg, Sweden
tommy.svensson@chalmers.se

Giorgio Calochira[1], Giuseppe Avino[2]
TIM Access Engineering Dept.[1], TIM Access Innovation Dept.[2], Turin, Italy
{giorgio.calochira, giuseppe.avino}@telecomitalia.it

Stefan Wänstedt, Pernilla Bergmark
Ericsson Research
Stockholm, Sweden
{stefan.wanstedt, pernilla.bergmark}@ericsson.com

Marja Matinmikko-Blue
Centre for Wireless Communications
University of Oulu
Oulu, Finland
marja.matinmikko@oulu.fi

*Abstract*— **From its conception, 6G is being designed with a particular focus on sustainability. The general philosophy of the H2020 Hexa-X project work on sustainability in 6G is based on two principles: to reduce direct negative life cycle impacts of 6G systems as much as possible (Sustainable 6G) and to analyze use cases that maximize positive environmental, social, and economic effects in other sectors of society (6G for Sustainability or its enablement effect). To apply this philosophy, Hexa-X is designing 6G with three sustainability objectives in mind: to enable the reduction of emissions in 6G-powered sectors of society, to reduce the total cost of ownership and to improve energy efficiency. This paper describes these objectives, their associated KPIs and quantitative targets, and the levers to reach them. Furthermore, to maximize the positive effects of 6G through the enablement effect, a link between 6G and the United Nations' Sustainable Development Goals (UN SDGs) framework is proposed and illustrated by Hexa-X use case families.**

## I. Introduction

Developing future networks towards 2030, there is a strong consensus among major stakeholders from industry and academia around the world that the network technology shall support and further accelerate the progress toward a better and more sustainable world [1, 2]. The most comprehensive, internationally elaborated framework for measuring this progress consists of the United Nations' Sustainable Development Goals (UN SDGs) [3]. To this end, the Hexa-X project, among its 27 use cases, proposes some which specifically address these UN SDGs and their underlying indicators [3]. This subset of use cases closely connected to the UN SDGs are put forward in the present paper.

Whatever its future use may be, future 6G systems must be designed and developed in a way that minimizes their direct negative life cycle impacts (Sustainable 6G) while at the same time specify capabilities that maximize their positive effects and suppress negative effects of different usages (6G for Sustainability). "Sustainable 6G" will help the Information and Communication Technology (ICT) sector stay in line with the prospective, normative trajectory set by the International Telecommunications Union (ITU) for the future greenhouse gas (GHG) emissions of the ICT sector to be compatible with the Paris Agreement [4]. "6G for Sustainability" will enable other sectors of society to reach sustainability targets, e. g. through 6G-induced substitution behaviors.

As a first step towards quantifying efforts towards a sustainability-embedded 6G, Hexa-X has proposed three project Key Performance Indicators (KPIs) regarding sustainability: i) reduction of emissions in 6G-powered sectors of society, ii) reduction of the total cost of ownership (TCO), and improvement of energy efficiency, by reducing the energy consumption per bit in networks. The first KPI pertains to "6G for Sustainability" while the second and third KPIs address "Sustainable 6G". Hexa-X also introduces targets for each KPI. The present article explains the rationale behind each KPI and target, as well as the methodology for the assessment of each KPI, and levers to reach aforementioned targets.

Finally, for 6G to follow through on these sustainability goals, novel architectural paradigms are introduced. They are centered on the principles of modularity, scalability and horizontality, and are detailed in this paper.

The structure of the present article is as follows. First, the various use case families proposed by Hexa-X are explored, and the way they will induce sustainability in various sectors of society is specified. Then, the sustainability KPIs and targets introduced by Hexa-X are detailed. Finally, the 6G architectural enablers that will allow reaching these targets are presented.

## II. Enabling sustainability by/for use cases

Future 6G systems should contribute to the sustainable development of society, and the set of use cases (UCs) introduced in the Hexa-X project illustrates how 6G could impact various sustainability aspects. The project defines six UC families, as depicted in Fig. 1. One of these UC families, "Enabling sustainability", is especially targeted towards sustainability-oriented development, while the other UC families contain some UCs that can contribute to meeting the UN SDGs in different ways. In any case, these UCs currently remain propositions, are open for debate and can change between the present time and the future introduction of

This work has been funded by the European Commission through the H2020 project Hexa-X (Grant Agreement no. 101015956).

commercial 6G. We now introduce some of the UC families, and how they match with sustainability objectives. For more details about the UCs proposed by Hexa-X, the reader is referred to [1, 2].

*B. Other UC families with sustainability-oriented UCs*

The "Telepresence", "Robots to cobots", "Massive Twinning" and "Trusted Embedded Networks" UC families

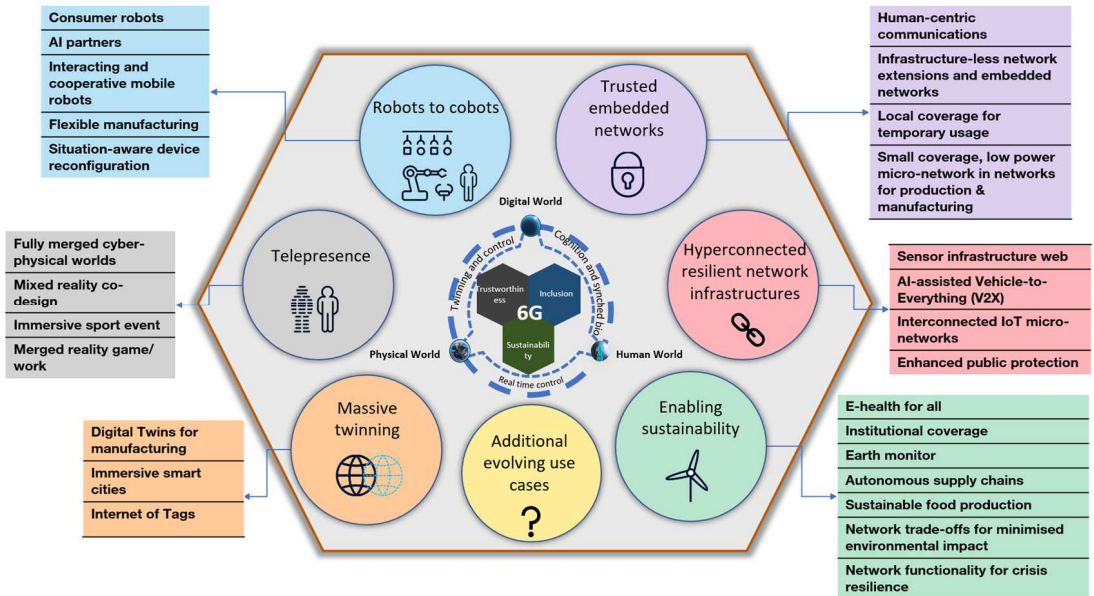

Fig. 1. The six use case families proposed by Hexa-X in 6G.

*A. The Enabling Sustainability UC family*

In the "Enabling Sustainability" UC family, several use cases are described which leverage 6G systems for different sustainability aims, such as minimization of environmental impacts, protecting the planet or facilitating inclusion. Some of the use cases rely on the provision of 6G connectivity everywhere, even in most rural or extreme rural areas, to deliver essential services such as consultation with a doctor, in remote areas lacking medical facilities (use case E-health for All). This use case could be combined with possible local sample taking and analysis, to complement consultations. This use case could contribute to UN SDG3 (Ensure Healthy Lives & Promote Wellbeing for All) and UN SDG12 (favor inclusion) [3].

Minimization of environmental impacts represent another strand for 6G for Sustainability. An example of use case is Autonomous Supply Chain, where 6G sensors or tags can optimize and automate the supply chain, allowing for an end-to-end tracking of goods, from fabrication to delivery, and then usage and recycling. This precise monitoring allows to optimize usage of resources and reducing waste.

Another direction or aim of 6G for Sustainability is the protection of the ecosystem. An illustrative use case is the Earth monitor. Involving deployment of sensors across the planet to monitor various aspects such as weather, climate change or biodiversity, such monitoring can allow to protect natural areas from disasters such as flooding and fires, or contribute to the protection of endangered species.

Food safety is another important goal for "6G for Sustainability". The use case family includes a use case labeled Sustainable Food Production, where massive twinning is an asset to monitor crops and guarantee the production of fruits or vegetables.

also introduce UCs with positive effect towards UN SDGs.

By enabling immersive experience, the Telepresence use case can offer solutions to avoid or limit travelling, potentially contributing to the reduction of emissions. Telepresence use cases, such as "Fully merged cyber-physical world", or "Mixed reality co-design" can potentially also contribute to societal sustainability by alleviating the feeling of exclusion, offering better quality of interactions for people in remote areas.

Other use cases can improve the efficiency of manufacturing processes, such as generalization of robots, as depicted in the Robots to Cobots use case family, and allow to optimize the usage of resources. In this use case family, the use case "Consumer robots" describes the multiplication of robots at home, performing multiple tasks, and such use case can contribute to social sustainability by supporting elderly or disabled people at home, thanks to the domestic robots performing for them hazardous chores and any action setting difficulties due to their conditions.

Massive Twinning use case family also includes use cases where twinning can enable fine control and monitoring of the usage of resources, and limit waste, such as in the case of smart city management (Immersive smart city use case"), for the management and regulation of the utilities (e.g., gas, heating) and different flows (e.g., traffic, transportation). In the "Trusted embedded networks" use case family, a use case such as "Human-centric communications" may improve health of people thanks to in-body sensors able to monitor some key parameters, and possibly perform adjustive measures if needed if anomalies are detected thanks to this continuous monitoring.

While a variety of UCs proposed by Hexa-X are expected to contribute to the fulfillment of the UN SDGs, it is important to recall that 6G, as all general-purpose technology, may have both positive and negative effects depending on its usage. Moreover, a use case which contributes to an SDG may have adverse effects on another. An important task for the project is hence to develop a technology that minimizes any unwanted side effects.

### III. HEXA-X SUSTAINABILITY KPIs AND TARGETS

Sustainability in 6G is being studied under three angles by Hexa-X: societal, economical, and environmental. As such, three objectives (one for each sustainability aspect) have been proposed. The societal objective is to reduce the emissions of other sectors through 6G-induced solutions. The economical objective is to reduce the TCO of 6G. The environmental objective pertains to the improvement of the energy efficiency of 6G. For each objective, the KPI, baseline, scope, methodology, quantitative target, and levers to reach such target (if applicable) are now presented.

*A. Societal target: enable the reduction of emissions of >30% CO2 eq in 6G-powered sectors of society*

or future situation. Since 6G is not deployed yet, we are doing the exercise of assessing a perspective, future situation, where both the 6G solution and possibly also the reference scenario are hypothetical.

In addition, the direct rebound effects also need to be taken into account, i.e., emissions associated with usage of a service which is not associated with modifying the baseline but occurring due to the convenience of the solution. Until now, both the baseline and consolidated detailed methods and standards that describe a rigorous methodology for evaluating the "enablement" impact of ICT on other sectors have been lacking [5].

In 2022, the ITU finalized a Recommandation to provide assessment methods for existing or defined solutions that became available in December 2022 [6]. This standard provides a basis for assessing the enablement effect but was developed with existing technologies in mind, assessments that could be based on measurements and actual usage of the ICT solution.

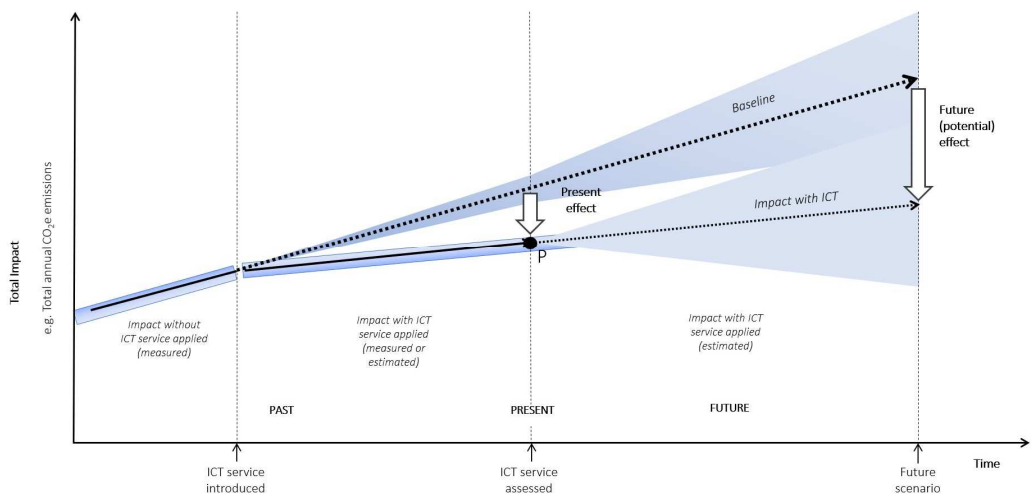

Fig. 2. Assessing the enablement effect [5].

The enablement effect (i.e., a positive second order effect) is commonly associated with solutions or services that could help reduce or avoid GHG emissions. The Hexa-X target calls for enabling reductions of emissions of >30% CO2 eq. in 6G-powered sectors of society. It is noted that the target specifically addresses positive effects, however the project is also acknowledging that 6G, as all general-purpose technologies, may also have use cases with negative effects.

Independent of expected outcome, a common approach for most work in this area is the definition of a baseline scenario without the solution, the definition of a scenario with the solution that reduces GHG emissions applied, and a comparison between the two. Fig. 2 illustrates a situation where a reference solution exists until the point of introduction of an ICT solution (in our case a 6G solution), the dotted baseline line then represents the evolution of the reference solution (i.e. the baseline or reference situation) if the 6G solution would not have been introduced, as compared to the lower line which estimates the situation with the 6G solution applied. Hence, the analysis is hypothetical, as the two scenarios cannot exist at the same time. The figure also introduces the importance of the time perspective and whether the assessment looks at a past, present

Hence, an important task of Hexa-X will be to explore how to enhance the standardized methodology to be applicable in the case of future technologies. For future technologies, proven case studies will be lacking, resulting in an inevitable additional level of uncertainty and the need to adapt any existing methodology. Moreover, the overall effect of 6G (the aggregated effect of all potential, future use cases) is beyond reach, as the total use of 6G cannot be foreseen.

Consequently, the evaluation of 6G can only be scenario-based, and refer to specific use cases, selected among those defined by Hexa-X: the current focus is on applications toward flexible/remote work enhancements and travel avoidance. The main challenges include the establishment of baselines, estimating impacts of future 6G solutions and their usage, and estimating the induced impact for a future scenario, potentially considering the direct rebound effect (not to mention the difficult extrapolation from case studies to larger populations). To achieve this target, the following items are under study in Hexa-X [2]:

**Baseline**: a non 6G powered service compared to a 6G-powered service. The main complexity is the modelling and data collection/estimation related to CO2 impact with and without 6G in a future setting.

**Methodology**: applying existing and developing methodologies to provide a transparent and well-founded result is always a challenging task associated with significant complexities and uncertainties, especially at this early stage of technology development. Moreover, systematic approaches are required to model the potential GHG reductions of enablement effects. Knowledge about technology itself is not sufficient to establish usage scenarios; strategies and policies may form cultural change and new personal and societal behaviors, and as such must be considered as well. In addition to taking into account the importance of such effects for the assessment, Hexa-X is currently studying how to consider not only technological aspects, but also the behavioral and cultural ones to identify the core aspects and actions, at both technical and organizational, behavioral, cultural levels that will be key to maximize the enablement effects and thus help reach the defined target. Without conducive actions, rebound effects may diminish and even, in some cases, overcompensate the potential advantages. However, it is acknowledged that rebound is a complex area and some rebound effects may actually amplify reduction effects [7].

**Scope**: Enablement effects should be defined and evaluated for specific use cases. Hexa-X is currently addressing the use cases defined in the project in order to identify the most suitable one(s) to be considered for the enablement effect analysis. Such analysis will be performed on a "what-if" basis, outlining the different scenario outcomes, should the selected use case(s) be made available and applied with an assumed effect or not. Moreover, in what measure a certain new technology will be adopted and used is also influenced by personal/societal culture, economic factors and behaviors. Importantly, enablement is calculated as a net effect after subtracting the footprint of the solution, unless that is deemed insignificant. This may imply some synergies with the economical and environmental objectives in terms of modelling.

*B. Economical target: reduce the Total Cost of Ownership of 6G by >30%*

A mobile operator's Total Cost of Ownership (TCO) for the introduction of a brand-new mobile system includes both Capital Expenses (one-time costs) and Operating Expenses (recurring costs), i.e., CapEx and OpEx, respectively [8]. The Hexa-X target calls for a reduction of the TCO for 6G by at least 30% with respect to current networks. In a typical mobile network today, CapEx is ~30% and OpEx is ~70% of the TCO over a 10-year period, with the Radio Access Network (RAN) being the biggest cost component in both CapEx (~50%) and OpEx (~65%) [9], then followed (from most to least impacting) by energy, backhaul, core network infrastructure, and other network costs (e.g., people, network management and maintenance, etc.). A breakdown of RAN CapEx shows that the largest cost components are site construction, spectrum, and equipment. Similarly, a breakdown of RAN OpEx shows that the largest contributors are power consumption, site rentals and operations.

To achieve the Hexa-X economical target, a methodology has been developed, in which the 6G TCO evaluation requires a baseline mobile network architecture to be properly identified. Such baseline architecture allows to assess the 6G TCO in relative terms (i.e., $x$% cost savings) with respect to it, by quantifying the potential cost reduction provided by the most promising 6G network enablers when deployed, for each of the cost items impacting the TCO. By considering that operators are currently deploying 5G networks based on both the new 5G Core network (5GC) and the NR (New Radio) access technology – i.e., the 5G NR Standalone (5G NR SA) – it is natural to assume the 5G NR SA as the baseline architecture for the 6G TCO evaluation.

For determining the cost structure and the share of each cost component (RAN, energy, and so on) in the overall 5G NR SA TCO, the study provided by the Global System for Mobile Communications Association (GSMA) has been considered [10]. Such work considers the dynamic interplay of a diverse mix of factors broadly falling into three groups: *cost drivers*, representing the "reasons why" a new (5G) network is needed, e.g., the mobile data traffic growth, the (operator-specific) strategy choices in terms of use cases being exploited for monetization, etc.; *cost accelerators*, that is, factors such as the RAN and the backhaul upgrades, the Edge Computing deployment, and so on, which increase the overall cost of owning and operating a (5G) network – and that can be classified as being CapEx or OpEx – as they are needed to cope with the presence of multiple cost drivers; and *cost optimisers*, which can serve as a catalyst to accelerate the (5G) network evolution while keeping the TCO at an affordable level from the operator's perspective. Typical cost optimisers include new RAN architectural approaches, e.g., virtual RAN (vRAN) instead of legacy distributed RAN (D-RAN), architectural enablers such as automation and Artificial Intelligence (AI) for planning and executing modern mobile network operations, low energy and or $CO_2$ reduction solutions such as liquid cooling replacing air conditioning for the equipment, and so on.

The final deliverable of Work Package 1 of Hexa-X will present a quantitative TCO analysis derived from the methodology described above. Specifically, the deliverable will assess the TCO reduction of a selected Hexa-X use case, i.e., the "Fully merged cyber- physical worlds" [1], which represents an application of a specific deployment strategy among the ones identified by the GSMA study, namely the "Rapid, full-scale 5G" deployment strategy. The analysis will provide a quantitative estimation on the potential cost savings achievable when the technological enablers identified by the project will be deployed.

*C. Environmental target: reduce energy transmitted per bit by >90%*

Several studies have shown that at each transition between two cellular generations, a reduction factor of 10 has been achieved in terms of energy consumption per transmited bit in wireless networks [2, 11]. This achievement has largely been obtained thanks to spectral efficiency induced by larger signal bands, hardware improvement in terms of integration, miniaturization, and processing, and substantial progress in sleep mode management systems.

Hexa-X addresses all the network segments including access, transport, and core networks. Our ambition is not only to consider networks but also to connect with services and content delivery points like cloud and data centers. The 5G NR was selected as a baseline, considering different traffic scenarios, including high or low data rates.

Measurement and assessment methods are now well-known and applied by mobile network operators. The assessment methodology is described in the European Telecommunications Standards Institute (ETSI) 203-228 standard which support operational networks. The evaluation method is the measurement of the energy consumed by a radio base station during a given time period (generally one hour) and the corresponding total traffic volume delivered by the base station to all the connected users. Ultimately, our goal is to match the infrastructure energy consumption to the level of the traffic load. The traffic volume strongly depends on signal bandwidth and frequency carriers. However, the energy consumption levers will essentially be related to the hardware capabilities and specifications that can be clustered into the following categories:

**Electronic components efficiency**. This category considers all the electronic layers that impact the global consumption including the baseband unit (computation part) as well as the radio unit (radio frequency (RF) Power Amplifier (PA)). The electronics component efficiency is the ratio between its DC power consumption and the quantity of transmitted/processed bits. The computation part was historically dependent on Moore's law and microchips integration. However, now-observed stagnation of Moore's law calls for new ways towards enhanced computing efficiency, such as co-designing of technological processes, architectures, and algorithms [12]. Meanwhile, the improvement of the RF amplifying part is driven by power amplifier technology improvements, and materials e.g., Gallium Arsenide in PA instead for its better performance at high frequencies.

**Bandwidth and signal characteristics.** This lever addresses the signal properties like frequency carriers, aggregation capabilities and bandwidth specification, in order to estimate the improvement of spectral efficiency.

**Artificial Intelligence and multi-goals optimization.** This new lever could bring very promising energy savings and optimizations while maintaining an equivalent quality of service (QoS). AI can be introduced to optimize sleeping periods of RF modules, as well as to adapt the needed resources to the user demand or other domains such as linearity and power amplifier optimization. Moreover, AI can also be used to detect energy consumption anomalies and overdimensioned sites that could be reenginereed to adapt the network resources to the targeted quality of service.

**Sleep-modes and network orchestration.** Sleep modes have been one of the main levers for decreasing the energy consumption of wireless networks this last decade. Their performance is closely related to the physical (PHY) and medium access control (MAC) layers design as well as to the signal characteristics. The main improvements have been achieved with the orthogonal frequency-division multiplexing (OFDM) structure which allows rapid sleep modes generally called micro-discontinuous transmission (micro-DTX). Also, the multiple input multiple output (MIMO) configuration now allows to switch off part of the antenna transceivers depending on the traffic demand. 6G PHY/MAC layers design should then natively consider sleep modes implementation to enhance their efficiency. New techniques such as lean carrier and deep sleep modes could then be implemented without a loss in QoS or user experience.

**Path loss reduction techniques.** Path loss in wireless communications is typically large. This loss comes from radio waves not reaching the intended receivers. The radio energy in these radio waves is thus wasted for communications purposes. In addition, radio waves reaching non-intended receivers will cause interference, thus degrading their capability to correctly receive signals intended for them. With precise beamforming, such losses can be substantially reduced, especially at mm-wave and (sub-)THz carrier frequencies where multi-path components are typically weak. Network densification is another enabler, since then the expected distance between transmitter and receiver can be reduced, and objects causing severe shadowing at mm-wave and (sub-)THz carrier frequencies can be mitigated thanks to macro-diversity. Combining densification with optimized sleep-modes and network orchestration has shown to enable overall energy savings [13]. Here, various distributed MIMO (D-MIMO) techniques [14], including integrated access and backhaul (IAB) [15], reconfigurable intelligent surfaces (RIS) and network-controlled repeaters (NCR) [16] all have a strong potential, but need further research to understand how to best optimize their use in various deployment and usage scenarios.

IV. ARCHITECTURAL COMPONENT FOR EFFICIENT NETWORKS

We now propose several enablers, needed to introduce a service-based architecture approach also for RAN. These enablers are from the start designed with the aim to make the network more efficient in general, and therefore more sustainable. For example, if signaling paths are shortened (fewer nodes involved) or the number of required messages is reduced, resource consumption is reduced.

The 3$^{rd}$ Generation Partnership Project (3GPP) defines a Service-Based Architecture (SBA), in which the control plane functionality and common data repositories of a 5G network are delivered by several interconnected Network Functions (NFs), each with authorization to access each other's services via standardized interfaces. RAN deployments have adopted some cloud-based enabler technologies that allow the virtualization of RAN functionality. With RAN virtualization and cloudification, it is today possible to use shared edge infrastructure for edge cloud deployments. Further cloudification of more RAN functionality that basically involves adapting SBA approach across 6G RAN-core network (CN) control planes becomes relevant. Thus, the development of 6G provides an opportunity to re-visit the roles of RAN and CN, taking full advantage of the latest cloud technologies in favour of a harmonized SBA that enables a more scalable and modular platform across RAN and CN. We now provide some ideas regarding how to plan for an SBA RAN.

First, dependencies between NFs should be avoided [17]. This includes dependencies between CN NFs and between CN NFs and RAN nodes. Further, in the process of creating independent NFs, we should avoid duplicated functionality, unnecessary options, and multiple processing points. We need to determine if there is any functionality that can be left out, i.e., a NF will still provide the necessary output/function but in a different way. The following steps will eventually show what functionality is crucial and will render parts of the network unusable when missing.

A first step in the process towards an efficient network is to structure functions in such a way that they can be analyzed, for the sake of identifying dependencies. Looking at 3GPP network functionalities, the first subdivision could be grouping non-user-equipment (UE) related and UE-specific functions. Next, we characterize the dependencies. Examples of dependencies are cross-function area dependencies and cross-NF dependencies within a function area. The prior can be dependencies between function areas such as user plane (UP) sessions, UE security, UE context management (/mobility) and UE-NF instance biding/message routing, all of them needed for e.g., Xn mobility. An example of dependency of the former kind is how a UP session may involve many different nodes, such as distributed unit (DU), central unit (CU)-CP, CU-UP, Access and Mobility Management Function (AMF), Session Management Function (SMF) and User Plane Function (UPF). There are other possible sources of dependencies, e.g., the ones originating from the evolution of network deployments, namely internal RAN split dependencies.

To overcome some of the identified hurdles for functional split dependencies, we need to change communication patterns. Instead of today's very long sequential procedures with several variants, we need small, independent, atomic transactions. Also, we need to remove unnecessary signaling proxy functionality, allowing direct communication between UE/RAN and CN functions. In other words, we reduce hierarchy and unnecessary dependencies. This can be done by keeping radio-related configurations together. Also, the CN – RAN separation can be maintained but communication needs to be made service-based with loosely coupled services. Also, it is necessary to optimize time-critical procedures, such as handover, Radio Resource Control (RRC) resume, radio link reconfiguration, and so on, without requiring tight bundling of services.

The 6G architecture shall be able to fully utilize the cloud platform. With a cloud-native network, it should be possible to streamline the RAN and CN architectures, i.e., to reduce some of the complexity. Taking full advantage of the cloud-native approach will ensure architectural and operational consistency across the RAN-Core and network management while minimizing potential backward compatibility issues.

All the above-mentioned functions need to be handled while at the same time considering sustainability, including circular economy principles, e.g., ensuring that the use of material for producing a network is optimized and that losses at end-of-life are minimized. Particularly, sustainability needs to be considered over the full life cycle. As the energy supply of networks increasingly becomes renewable and low carbon, the importance of materials and production processes is expected to increase.

V. CONCLUSION AND FUTURE DIRECTIONS

In this paper, we first reviewed a selection of use cases proposed by Hexa-X and linked them with the UN SDGs. Then, we described the societal, economical, and environmental sustainability KPIs and associated targets put forward by Hexa-X, underlying the two principles of sustainability in 6G development: Sustainable 6G and 6G for Sustainability. A major result of this paper is the definition and methodology considerations for the Hexa-X 6G sustainability KPIs. Finally, we presented some architectural enablers developed to reach the sustainability targets.

In future works, Hexa-X will focus on the methodology to measure the sustainability KPIs, as well as on enablers, with the goal to feed back to and be anchored with 6G technology development. We will continue to assert the role of the Hexa-X technology and use cases in effectively making 6G a core asset for society's sustainability.